\documentclass[12pt,preprint]{aastex}
\usepackage{color}
\usepackage{graphicx}

\begin{document}

\def\NB#1{[NB: {\it #1}]}
\def\DK#1{[DK: {\it #1}]}   
\def\DT#1{[DT: {\it #1}]}   
\def\JO#1{[JO: {\it #1}]}   
\def\NS#1{[NS: {\it #1}]}   

\def \etal {{\it et~al.}}
\def \nhi  {$N_{HI}$}
\def \ndi  {$N_{DI}$}
\def \lnhi  {$Log~N_{HI}$}
\def \lndi  {$Log~N_{DI}$}
\def \nhe  {$N_{HeII}$}
\def \qone {PKS~1937--1009}
\def \qtwo {Q1009+2956}
\def \qthree {Q0130--4021}
\def \qfour {HS~0105+1619}
\def \qfive {QSO 1243+3047}
\def \qhst {PG~1718+4807}

\def \ETA {$\eta $}
\def \het {$^3$He}
\def \hef {$^4$He}
\def \lisv {$^7$Li}
\def \yp      {Y$_p$}

\def \btemp {$b_{temp}$}
\def \bturb {$b_{turb}$}
\def \ly    {Lyman}
\def \mf    {$\times 10^{-5}$}

\def \ob {$\Omega _b$}


\newcommand{\kms}{km s$^{-1}$}
\newcommand{\cmm}{cm$^{-2}$}
\newcommand{\cmmm}{cm$^{-3}$}
\newcommand{\escsh}{$\mathrm{ergs}
        \mathrm{\ s}^{-1}
        \mathrm{\ cm}^{-2}
        \mathrm{\ sr}^{-1}
        \mathrm{\ Hz}^{-1}$}

\newcommand{\chisquared}{$\chi^{2}$}

\newcommand{\omegatotal}{$\Omega _{total}$}
\newcommand{\omegam}{$\Omega _m$}
\newcommand{\omegalambda}{$\Omega _{\Lambda}$}
\newcommand{\omegabaryon}{$\Omega _b$}
\newcommand{\omegabaryonhsquared}{$\Omega _bh^{2}$}
\newcommand{\hubbleconstant}{$H_{0}$}
\newcommand{\lcdm}{$\Lambda$CDM}
\newcommand{\ocdm}{OCDM}

\newcommand{\pk} {$P(k)$}

\newcommand{\lya}{Ly$\alpha$}
\newcommand{\lyb}{Ly$\beta$}
\newcommand{\lyg}{Ly$\gamma$}
\newcommand{\lyd}{Ly$\delta$}
\newcommand{\lyaf} {Lyman-$\alpha$ forest}
\newcommand{\zabs}{$z_{abs}$}
\newcommand{\zem}{$z_{em}$}
\newcommand{\da}{$D_A$}

\newcommand{\jtwentyone} {$J_{21}$}
\newcommand{\jzero}{$J_0$}
\newcommand{\jnu}{$J_{912}$}
\newcommand{\jovi}{$J_{197??}$}

\title{
Relative Flux Calibration of Keck HIRES Echelle Spectra\altaffilmark{1}} 

\author{ 
Nao Suzuki\altaffilmark{2},
David Tytler\altaffilmark{3},
David Kirkman,
John M. O'Meara,
\& Dan Lubin \\
Center for Astrophysics and Space Sciences;\\
University of California, San Diego; \\
MS 0424; La Jolla; CA 92093-0424\\}

\altaffiltext{1} {Based on data obtained with the Kast spectrograph 
on the Lick Observatory 3-m Shane telescope and with the HIRES and 
ESI spectrographs
at the W.M. Keck Observatory that
is a joint facility of the University of California, the California
Institute of Technology and NASA.}
\altaffiltext{2} {E-mail: suzuki@ucsd.edu}
\altaffiltext{3} {E-mail: tytler@ucsd.edu}

\begin{abstract}

We describe a new method to calibrate the relative flux levels in
spectra from the HIRES echelle spectrograph on the Keck-I telescope.
Standard data reduction techniques that transfer the instrument response
between HIRES integrations leave errors in the flux of 5 -- 10\%, because
the effective response varies. The flux errors are most severe near
the ends of each spectral order, where there can be discontinuous jumps.  
The source of these errors is uncertain, but may include changes in the
vignetting connected to the optical alignment. 
Our new flux calibration method uses a calibrated
reference spectrum of each target to calibrate individual HIRES
integrations. We determine the instrument response independently for
each integration, and hence we avoid the need to transfer the
instrument response between HIRES integrations.  The procedure can be
applied to any HIRES spectrum, or any other spectrum. While the 
accuracy of the method depends upon many factors, we have been able to
flux calibrate a HIRES spectrum to 1\% over scales of 200~\AA\ that include 
order joins. We illustrate the method with spectra of \qfive\ towards which 
we have measured the deuterium to hydrogen abundance ratio.


\end{abstract}

\section{Introduction}

In recent decades, the combination of large aperture telescopes and
high resolution spectrographs have allowed for precision analysis of a
variety of astrophysical objects.  Echelle spectrographs are the
instrument of choice for high resolution, and most large telescopes
now have one \citep{vogt87, diego90, dekker00, noguchi98, tull98,
mclean98}.

Echelle gratings can give spectra with high spectral resolution, with
a large slit width, and a large wavelength range in a single setting
\citep*{schroeder87}.
An echelle grating disperses the spectrum into many tens of spectral
orders, which are then cross dispersed by a second dispersive element
so that the orders can be placed, one above the other, on a
rectangular CCD detector.  It is difficult to combine the spectra from
the many spectral orders of an echelle to produce a single continuous
spectrum.  This difficulty arises because the response varies rapidly 
across each
order, and at a given wavelength is usually different in different
orders.



High quality relative flux calibration of echelle spectra is highly
desirable in many scientific applications.  For example, accurate flux
levels over a large range of wavelengths makes it much easier to place
continuum levels on spectra with pervasive blended absorption, such as
the Lyman alpha forest absorption seen in high redshift QSO spectra.



In this paper, we discuss the relative flux calibration of spectra
from the HIRES echelle spectrograph on the Keck-I telescope
\citep{vogt94}.  We do not discuss absolute flux
calibration, as it requires additional calibration data and
is not necessary for our absorption line work.
We intend that a spectrum with relative flux calibration has the correct
shape over some range of wavelengths, and differs from an absolute calibration
in only the normalization.

We developed the methods described in this paper to improve our
measurements of the primordial Deuterium to Hydrogen abundance ratio
in QSO absorption systems, for which the bulk of our spectra come from
HIRES.  We have found that we obtain more accurate and reliable
estimates of the absorption column densities when we use
spectra with accurate relative flux.  High quality flux calibration
was not a major design goal for HIRES, and we have found that special
steps must be taken to obtain the quality of calibration that we need
for our work on D/H.  The usual
\citep{iraf_echelle,iraf_slit,echelle_starlink}
methods of echelle flux calibration appear inadequate
for reasons that we do not fully understand.  This inadequacy
motivated the development of the methods we describe.

We would like to both minimize the flux errors in our spectra, and to
estimate the size of errors which remain after our calibrations.
We shall find that the flux calibration errors depend on wavelength and
they are correlated on various wavelength scales. We would like to estimate 
the size of the errors on these different scales.
We would also like to make the error in the relative flux calibration 
less than the photon noise on some relevant scale. For example, when we 
fit a flat
continuum level to a 50 pixel segment of a HIRES spectrum with signal to 
noise ratio of 100 per pixel, the photon noise on the continuum level is 0.14\%.

We found that is harder to approach a given accuracy in relative flux
calibration in many places.  These places include the regions
where echelle orders join, regions where spectra have lower signal to noise
ratio, 
wavelengths in the near UV, and in general as the wavelength range increases.  
It is very hard to get flux calibration errors of $<1$\% over 
even a wavelength range of $<40$~\AA\ within one HIRES order. 
Fortunately for our absorption line work, errors that vary
smoothly over wavelength scales $> 40$ \AA\ are not as serious as
smaller scale errors.

In this paper, we describe a way of calibrating the relative flux in a
HIRES spectrum using well calibrated reference spectra of the same
target to transfer flux information from other spectrographs to the
HIRES spectra.  We force each HIRES spectrum to have the shape of
these reference spectra.  This method should correct a variety of flux
calibration errors, both from variable vignetting and differences in
spectrum extraction and reduction.  The method could also be applied
to spectra from various spectrographs, and not just echelles.  This
method is based on that introduced by \citet*{burles97,burles98b}
and applied with improvements to \qfour\ in \citet{omeara01}.

The paper is organized as follows: we first describe the nature of
inconsistencies between repeated HIRES observations of a target, and
how this impacts flux calibration.  Second, we describe the spectra we
used to illustrate our flux calibration method. Third, we describe at
first in overview, and then in detail, our methods for flux
calibration.  Fourth, we describe how we combine the HIRES orders that
we have flux calibrated individually. Finally, we discuss the accuracy
and errors of our method.

\section{Description of the HIRES flux calibration problem}

\begin{figure}
\epsscale{0.9}
\plottwo{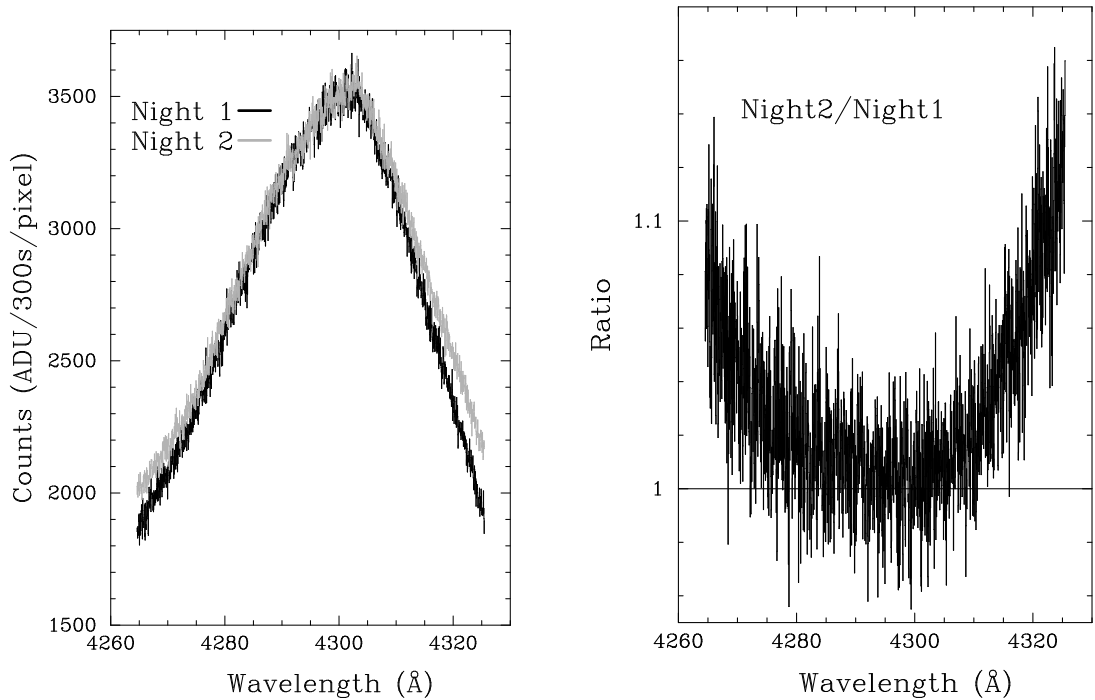}{x3v2.ps}
\caption{
Two 300 second integrations of the star G191-B2B taken with near identical
HIRES setups on consecutive nights: 19 (solid) and 20
(light) September, 2000.  The right panel shows the ratio of the two.
These spectra were taken with the C5 dekker (1.14 arcsec slit), and
the image rotator was set to align the sky vertical along the slit.
The September 19 integration had sidereal time ST = 05:16:48 hours,
telescope elevation EL = 56.96 degrees, and image rotator physical
angle, as measured looking at the prism, of IROT2ANG = 195.9357
degrees. For September 20th we had ST = 05:02:24, EL = 57.03 degrees
and IROT2ANG = 195.8646.  The September 20th spectrum was multiplied
by 1.06 to give similar counts to the other near 4300~\AA\ to correct
for differences in the atmospheric transparency, seeing and the loss
of light at the slit.
 \label{X2}  }
\end{figure}

The usual methods of flux calibration appear inadequate when applied to
HIRES spectra.  When we determine the instrument response by observing
a spectrophotometric standard star, different exposures of the same
star give different instrument response functions, even when we
believe that the exposures were taken with the same instrument
configuration.  For example, in Figure \ref{X2}, we show the signal
extracted from two HIRES integrations of G191-B2B.  The exposures were
taken on consecutive nights, with similar instrument configurations.
On the second night the star was 0.07 degrees higher in the sky and
the image rotator (Appendix \ref{rotator}) physical angle differed by
0.071 degrees to compensate for the change in parallactic angle.  The
spectra were both extracted with Tom Barlow's MAKEE software (April
2001 version). We show the ``raw" ADU from the CCD before division by
the flat field integration or any other calibration.

\begin{figure}
\includegraphics[angle=270,scale=0.45]{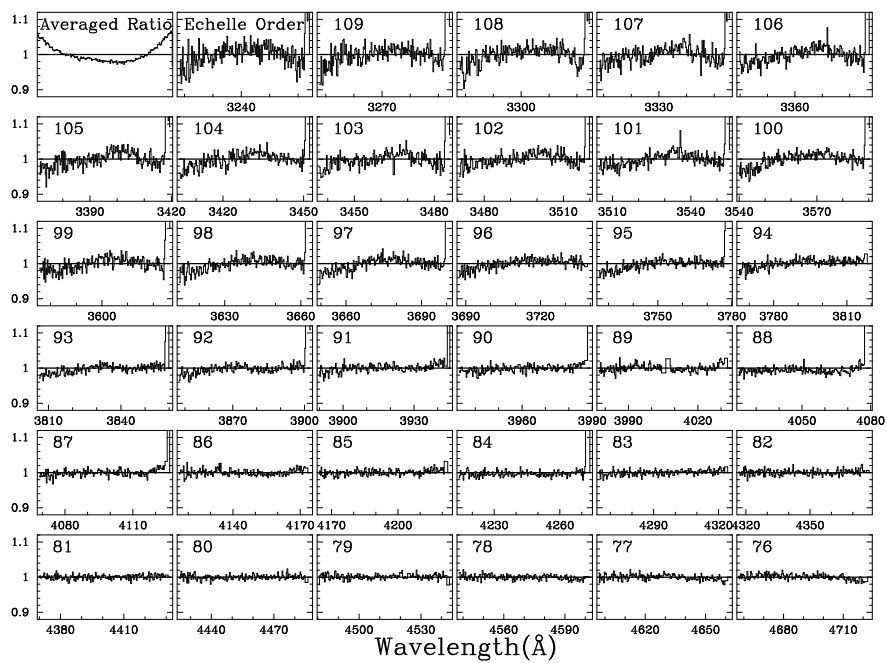}
\caption{
The ratio of the two spectra shown in Fig. \ref{X2}, divided by
average ratio from the ten orders with the largest wavelengths. We show this
average in the top left panel, and other panels each show one HIRES order.
We show the full amount of each order
that fell on the CCD, which includes some wavelength overlap between
adjacent orders. We have re-binned the original
2048 pixels into 205 pixels for presentation.
\label{X22c}  }
\end{figure}

The differences between the two exposures shown in Figure \ref{X2} are
large and unexpected.  In particular, even though the exposures were
taken on different nights, we did not expect to see large ($\sim
10$\%) differences within a single order, even after the two exposures
were normalized to have the same mean flux in that order.
The differences are largest ($\sim$ 10\%) at the ends of the order.
Similar differences (both in shape and magnitude) are present in each
observed spectral order.  However, the form of the difference is not
precisely the same for each order, as demonstrated in Figure
\ref{X22c}.  \citet{barlow97} also noted the possibility of systematic
flux errors of 10\% near order overlaps.

Flux differences have also been reported for the Subaru telescope
HDS echelle spectrograph
that sometimes shows 10\% changes during observations \citep{aoki02}.

We have examined approximately 20 other pairs of standard star integrations 
from HIRES that have similar instrumental setups.
Most show differences of order 10\%, though the exact shape of the
differences is not always the same.  In addition to the ``U" shaped
ratios (seen in Figure \ref{X2}), we see three other shapes for
ratios: near flat ratios, tilted and ogive (or ``fallen S") shapes.  In
each case, the shapes of the ratios vary gradually order-to-order in a
systematic way, such that adjacent orders have similar shapes. The
shapes of the ratios vary much more between pairs of integrations than
they do from order-to-order for a given pair of integrations.

Approximately 30\% of integration pairs give flat ratios that indicate
that the instrumental response was very similar for the two
integrations, which should make flux calibration simple.  A cursory
examination did not show any predictors (e.g. telescope elevation,
rotator angle, position of target on the slit, seeing) as to which
pairs would be similar and which different.

\begin{figure}
\epsscale{0.9}
\plottwo{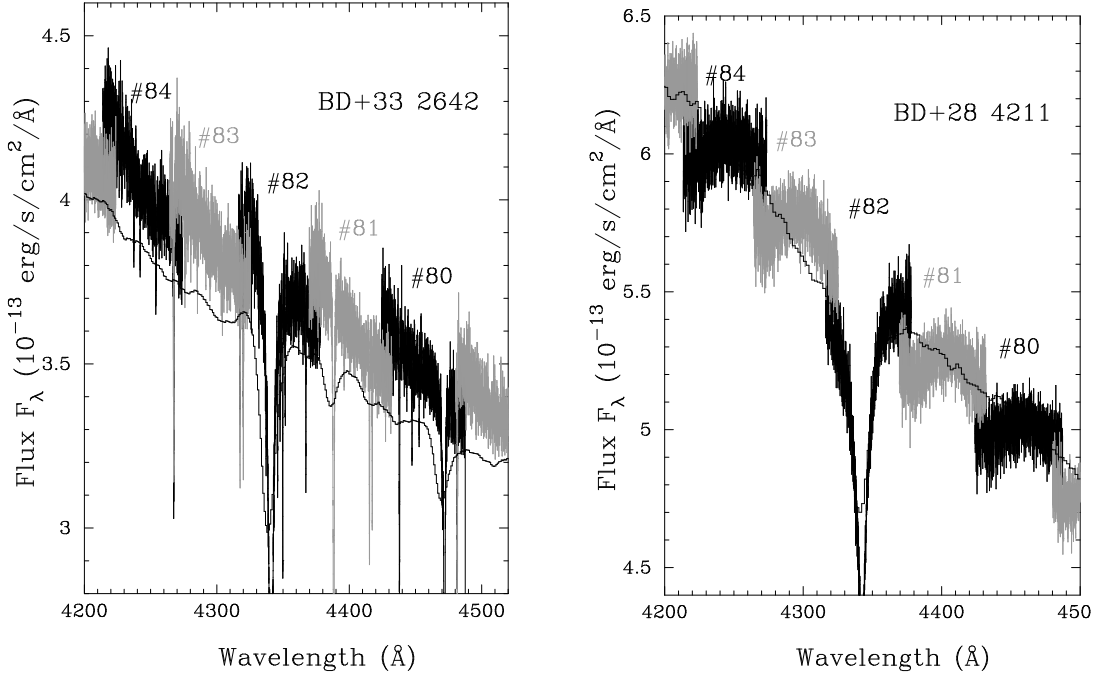}{x16bv5.ps}
\caption{
HIRES spectra of stars that we flux calibrated in the usual manner.
In both cases we see large flux calibration errors. In other cases we
can obtain smaller errors.
Left, 
HIRES integration of star BD+33 2642.
We determined the response function of HIRES by
comparing a HIRES spectrum of G191-B2B, taken on the same night
(September 19, 2000), to a model spectrum with known flux (Appendix
\ref{appa}).  The continuous line shows a lower resolution HST FOS
spectrum of BD+33~2642. We may adjust the HIRES spectrum vertically to
account for slit losses.  The right panel is the same, but for a HIRES
integration of BD+28~4211 obtained October 10, 1999, calibrated with a
HIRES spectrum of G191-B2B, taken on the same night. The continuous line
is a STIS spectrum of BD+28~4211.  \label{X16} }
\end{figure}

In Figure \ref{X16}, we show HIRES integrations on two stars that we
flux calibrated in the usual manner, each using a response function
determined from a HIRES exposure of a standard star.
The spectra that we calibrated differ from the known flux levels by large
amounts over a wide range of wavelengths. The main deviations are
systematic across each HIRES order.  We also see large differences at
the order overlap wavelengths.  Different ways of combining spectral
orders leave different flux calibration errors. If we take the
mean of the signals then the flux will jump in a single pixel, by up
to 10\%, where each order begins and ends.  There will be
approximately 70 such jumps in a complete HIRES spectrum with 36
orders.

In the past, we have attempted to reduce the flux errors
by fitting continua to each order, and dividing by these continua,
before we take the mean flux in a wavelength overlap. This method is
unsatisfactory because it is very hard to ensure that the continua
that we fit to adjacent orders have both the same flux levels and the
same shapes, and we loose flux information.

For some unknown reason, long integrations on a QSO show
smaller differences than short integrations on a bright star. A dependence on
integration time might relate to some averaging, perhaps related to
the target position on the slit. A dependence on brightness might
relate to the 100 times lower signal in the QSO integrations, and
perhaps subsequent differences in the spectrum extraction.  We still
have difficulty flux calibrating QSO exposures, even though they
appear to be more stable than exposures on standard stars, because usual
procedures still require that we determine the instrument response from a
standard star exposure.

The variations we see in standard star exposures indicate that there
is some instability in either Keck+HIRES, in our data reduction
processes, or in both.  We have investigated two possible origins for the
instability: variable vignetting inside of HIRES, and inadequate the
extraction of the spectra. We do not see a clear signature of either
in our spectra, but we do know that the vignetting is expected to change.

We have explored the possibility of
extraction errors by varying the type of profile used during
extraction, and the profile width, but found no differences from the
standard extraction results from MAKEE.
We also measured flux ratios from the raw counts in the CCD images that were
similar to those in the spectra extracted by MAKEE.

We know that the vignetting inside HIRES changes with the position
angle of the sky image on the HIRES slit and with the telescope
elevation (Appendix \ref{rotator}).  
When the image rotator is used there are two main options.
We can use the image rotator in ``Vertical angle mode" to keep the
vertical angle along the slit, so that the position angle varies with
the position of the target in the sky.  The rotator can also be used
to keep a desired position angle along the slit.  If the rotator is
not used, the relevant angle is the telescope elevation.  The
variation in vignetting arises from a known mechanical and optical
misalignment between the Keck-I telescope and HIRES, and the expected
amount of change in the vignetting, from ray tracing kindly provided
by Steve Vogt, is approximately 10\%, consistent with our
observations.

We can account for why the ratios of HIRES spectra have similar shapes
across all orders if the variable vignetting happens after the
echelle, and before leaving the cross-disperser. In this part of HIRES
the light from the red end of each order is separated from that in the
blue end, but all orders are coincident. We might explain the shape of
the ratios, and the similarity from order-to-order, if a varying
amount of light misses the top and bottom of the cross-disperser (the
grooves are vertical), where the red and blue ends of the orders land.
This variation could arise when the cone of light from the telescope
tips up and down in the vertical plane that connects the center of the
tertiary mirror and the HIRES slit.

However, we suspect that vignetting is not the sole cause of the
variations in the flux, because we have seen variations in spectra
taken under apparently identical instrumental conditions (same
elevation, image rotator setting, and target location on the slit) on
consecutive nights, as we saw in Figure \ref{X2}.

The changes might also come from differences in the extraction of the
spectra from the CCD image, e.g. if we do not extract a fixed
proportion of the flux recorded at each wavelength.  Differences in
the extraction of spectra, including multiple integrations on a given
target, are likely whenever there
are changes in conditions, such as the location of the target along
the slit, the seeing, the sky brightness and the amount of signal
recorded.  However, extraction problems seem an unlikely explanation
for standard stars that have high signal to noise ratio.

\section{Spectra we will use to Illustrate our Method}

Here, we introduce the spectra that we will use to illustrate our
method of relative flux calibration. This is the set of spectra that
we used to measure D/H towards \qfive\ ($z_{em} = 2.64$, V=16.9;
Kirkman \etal\ 2003).  For our D/H work, we were mostly interested
the flux calibration in a 40~\AA\ region centered on the damped \lya\
line near 4285~\AA, and on the Lyman limit near 3210~\AA .  We began
the development of the methods using a similar set of spectra of
\qfour\ that we had used to make an earlier D/H measurement
\citep{omeara01}.

We used 5 spectra from the Kast spectrograph on the Lick 3-m
telescope, and one from the ESI echelle spectrograph on the Keck-II
telescope.  We used both the Kast and ESI spectra separately to make
independent flux calibrations of 8 integrations from HIRES.  Further
details of the observations, including the dates, the resolution, the
mean S/N, and plots of the spectra are in \citet{kirkman03}.  All
of the spectra that we used were shifted into the heliocentric frame,
and placed on a logarithmic wavelength scale with a constant velocity
increment per pixel, although with different increments for different
spectra.

\subsection{Spectra from Kast}

The Kast double spectrograph uses a
beam splitter to record blue and red spectra simultaneously in two
cameras.  For \qfive\ we have five KAST integrations, one from 1997,
and two each from 1999 and 2001. All integrations were obtained using
the d46 dichroic that splits the spectrum near 4600~\AA, the 830
line/mm grism blazed at 3460~\AA\ for the blue side, and the 1200
line/mm grating blazed at 5000~\AA\ for the red side.  We reduced all
exposures with the IRAF package {\tt longslit}.

\subsection{Spectra from ESI}

ESI covers from 3900 -- 11,000~\AA\ in ten
overlapping orders \citep{epps98,bigelow98,sheinis00}.
We have one exposure of \qfive\ using a 1'' slit, taken in the
echellette mode on January 11, 2000.  From the same night, we also
have an exposure of the flux standard star Feige 110.

\subsection{Spectra from HIRES}

Our HIRES spectra of \qfive\ all used similar instrumental setups.
The angle of the HIRES echelle was the same for all exposures, and
placed the center of each order near the center of the CCD. The
cross-disperser angle was also similar for all exposures, except for
one exposure that extended to much larger wavelengths.  The image
rotator (Appendix \ref{rotator}) was used in vertical mode, to
minimize slit losses from atmospheric dispersion.  We used the C5
decker, which provides an entrance aperture to the spectrograph with
dimensions $1.15^{''} \times 7.5^{''}$. In each case we placed the
target near the middle of the slit. The spectra were all recorded on
the engineering grade Tektronix CCD with 2048~$\times$~2048 pixels
that has been used in HIRES since 1994.
The HIRES pixel size is 2.1~\kms .

The CCD is large enough to record beyond the free spectral range for
all orders at wavelengths $< 5200$~\AA, and we placed the spectra on
the CCD such that we did record this flux for all such orders.  All
but one integration covered from near 3200\AA\ out to 4700\AA\ in
approximately 36 orders.  These integrations were 7200 -- 9000 seconds
long.  The S/N in each integration is near 3 per 2.1~\kms\ pixel near
the Lyman limit at $\sim$ 3200~\AA , and rises to near 35 at the peak of
the \lya\ emission line at $\sim$ 4400~\AA .

The HIRES spectra we use are the normal output from the MAKEE
software, which are the raw counts spectrum divided by spectra
extracted from flat field integrations.  In addition, the CCD defects
were marked with negative error values.  These spectra differ from the
raw counts that we discussed in the previous section in that the flat
field division has removed most of the variation across the orders due
to the blaze and vignetting. Although this flat field division may
introduce additional undesirable changes in the relative flux, we
proceed with these spectra because it is imperative that the CCD
defects have been removed.

\subsubsection{HIRES Spectral Resolution}

We measured the instrument resolution of HIRES by fitting Gaussian
functions to narrow, apparently un-blended lines in a single
Thorium-Argon arc integration taken before one of the QSO
integrations.  We fit hundreds of arc lines in all parts of the
spectrum, and found a dispersion of $\sigma = 3.4 \pm 0.1$~\kms, which
corresponds to a FWHM spectral resolution of $8.0 \pm 0.2$~\kms
($b_{ins} = \sqrt{2}\sigma = 4.81 \pm 0.14$~\kms).  We did not detect
any variation in the resolution with wavelength.  However, we did
detect that the arc lines are not Gaussian in shape, with more
extended wings, such that the best fitting $\sigma$ increases when we
extend the fitting range around an individual line.  We also expect
that the spectra will have slightly different FWHM from the arc
spectra, because the illumination of the slit is different.
The wavelength shifts that we describe in the next section suggest that the
resolution will depend in part on the guiding and seeing.

\subsubsection{HIRES Wavelength Shifts}
\label{hireswave}

\begin{figure}
\includegraphics[angle=270,scale=0.45]{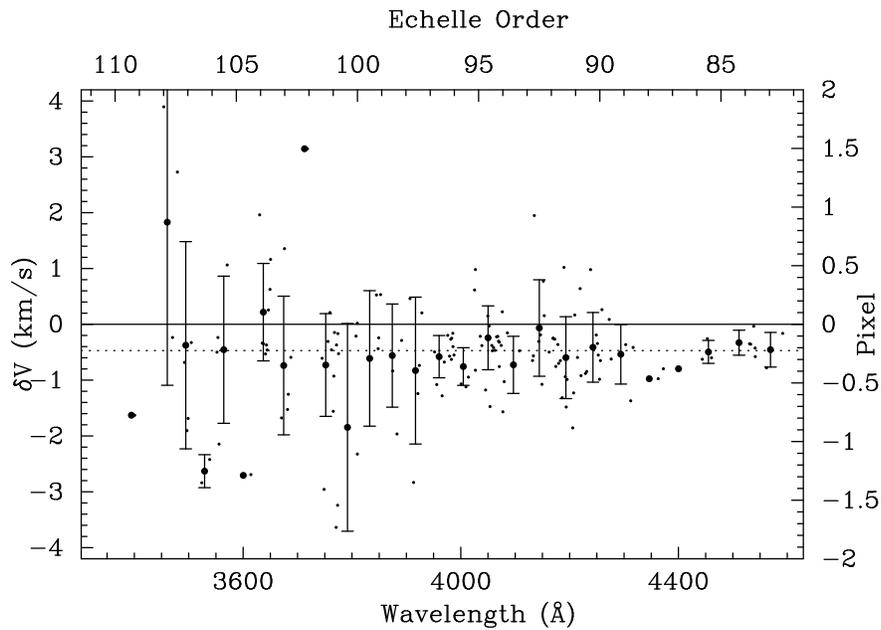}  
\caption{
The wavelength shifts measured between two consecutive HIRES integrations
for \qfive , both taken on March 13, 2000.  The small points show
shifts measured by cross-correlating blocks of approximately 20 pixels that 
contain an absorption line.  The larger points show the
mean shift per order, and the vertical bars show $\pm 1\sigma $ from
the distribution of the measurements in that order. The dotted line
shows the mean shift between the two integrations.
\label{X20a}  }
\end{figure}

We measured wavelength shifts between the HIRES integrations and we
shifted the spectra onto the same scale to correct these shifts.  We
measured the cross-correlations between each of the 7 integrations
with the eighth that we used as the reference. An example of the
shifts is shown in
Figure \ref{X20a}. Comparisons of other pairs of integrations often
show a much larger dispersion in the measured shifts.  
In all cases, the shifts measured in each order are consistent with a
single shift for each integration.  The shifts had a standard
deviation of 0.7 \kms, which is 30\% of one HIRES CCD pixel.  Each
shift was measured to an accuracy of 0.13~\kms , which we determined
from the scatter in the shifts that we measured separately for each
order.

These shifts may arise from differences in the placement of the QSO
light in the HIRES slit, which projects to approximately 4 HIRES
pixels.  The 0.7 \kms\ rms shift is approximately 9\% of the FWHM
resolution, which itself is similar to the slit width.  The shifts do
not correlate with hour angle or the correction that was applied for 
the Earth's orbital motion ($<30$~\kms ) and spin ($<0.4$~\kms ).  The
shifts are
also larger than we expect from wavelength scale errors.  However, we
did find much larger wavelength scale errors when we did not 
ensure that MAKEE used enough arc lines to determine the dispersion
solution for each order.

\section{Overview of the Method}

There are three main steps in our procedure to apply relative flux
calibration to HIRES spectra.

\begin{itemize}
\item Step 1: Flux calibrate a high quality reference spectrum of
      the target.  
\item Step 2: Flux calibrate the HIRES echelle orders with the
      reference spectrum.  This imposes the long scale ($> 10$ \AA)
      spectral shape of the reference spectrum upon each HIRES echelle
      order.  The flux information on smaller scales (e.g. absorption
      line profiles) still comes from HIRES.
\item Step 3: Combine the HIRES orders. We first sum the integrations
      and then join the orders to give one continuous spectrum.
\end{itemize}

These procedures do not replace standard CCD spectrum extraction
procedures. Rather, they should be thought of as a ``software patch",
applied to correct errors that remain after the spectra have been
extracted, and perhaps calibrated, in the usual way.  Reasonably well
calibrated HIRES spectra are required as inputs to our method, because
the flux information on scales $< 10$ \AA\ will still come directly
from the HIRES spectra.  The procedure
is not unique, and we expect that other sequences might be appropriate
for different spectra.

We now discuss in further detail each of the steps outlined above.

\section{Step 1: Flux Calibrating the Reference Spectra of the Target}

Since we can not transfer information about the instrument response
between exposures, we must ``self-calibrate'' each exposure we take
with HIRES.  The first step of our method is thus to obtain a well
calibrated spectrum of the target.  The reference spectrum must
come from a well calibrated and stable spectrograph.  In our work on
Q1243+3047, we obtained reference spectra from both Lick+KAST and
Keck+ESI, and compared them to gauge the accuracy of the final flux
calibration.

\subsection{Flux Calibration of Kast Spectra}

\begin{figure}
\epsscale{0.6}
\plotone{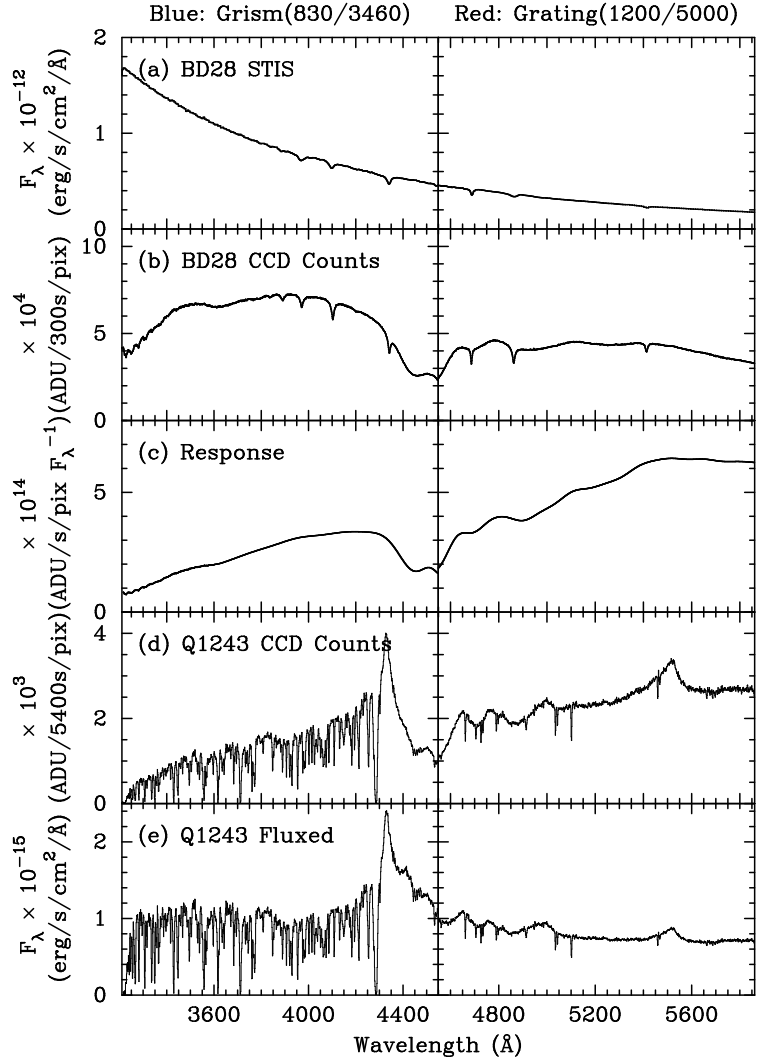}
\caption{
Steps in the calibration of the flux in a Kast spectrum.
Panel (a) shows the STIS spectrum of star BD+28 4211. Panel (b) shows the
raw counts recorded in two simultaneous Kast integrations,
one with the blue camera (left) and one with the red camera (right).  We do
not show the wavelength overlap on
either side of the central wavelength of the dichroic beam splitter (called
d46), which we show with the vertical line near 4450~\AA .    
We had moved the ``x-stage" that holds the CCD of the blue camera of Kast
to sample wavelengths well beyond the peak of the \lya\ emission line.
Panel (c) shows the response function (b)/(a), and panel (d) shows the raw
counts in one 5400 second integration on \qfive .
Panel (e) shows the flux calibrated spectrum, (d)/(c).
\label{X9}  }
\end{figure}

To calibrate the flux in our Kast spectra of \qfive , we took flux
information from a model spectrum of G191~B2B for the 1997 spectrum,
and a STIS spectrum of BD~28~4211 for the 1999 and 2001 exposures.  We
discuss the reasons behind our choice of these standard stars in
Appendix A.  We measured and matched the resolution of the observed
and reference spectra of the standard stars before we used them to
calculate the Kast response function.  The dip in the response at the
end of blue CCD and the beginning of red the CCD is due to the
changing efficiency of the beam splitter.  In Figure \ref{X9}, we
illustrate the flux calibration process.

The spectrum shows atmospheric ozone absorption lines
\citep{schachter91} as the wiggles of the raw CCD counts (panel b)
below 3400~\AA .  Their strength depends on the temperature of the
ozone layer and the effective airmass of the integration, and we have
not made appropriate adjustments.  We left the wiggles un-smoothed in
the response (Figure \ref{X9}, panel d) to help partially remove their
effect from the quasar spectrum.

\begin{figure}
\includegraphics[angle=270,scale=0.45]{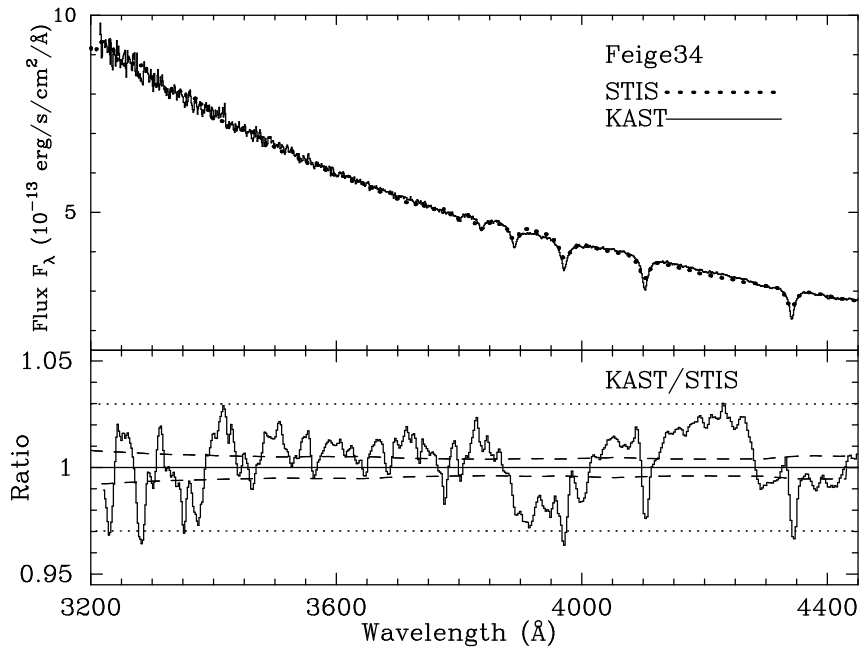}
\caption{
Top Panel: Two spectra of the star Feige~34, one from STIS (solid line,
\citealt{bohlin01}), and
the other a Kast spectrum that we have flux calibrated (dotted).
We calibrated the Kast spectrum with Kast spectrum of the star BD+28~4211.
We also normalized the Kast spectrum to have the same mean flux as the STIS 
spectrum to adjust for slit losses.
Bottom Panel: Ratio of the two spectra in the top panel.
The dotted line is the error for the STIS spectrum (approximately 1\% random 
and 3\% systematic), and the dashed line is the
random error from the photon noise in the Kast spectrum.
These errors are for 2~\AA\ pixels.
\label{X23}  }
\end{figure}

In Figure \ref{X23}, we show the accuracy of the flux calibration of a
Kast spectrum of another star.  The flux residuals between our
calibrated KAST spectrum and a STIS spectrum of the same star are less
than 3\%.

\subsection{Flux Calibration of ESI Spectrum}

\begin{figure}
\epsscale{0.6}
\plotone{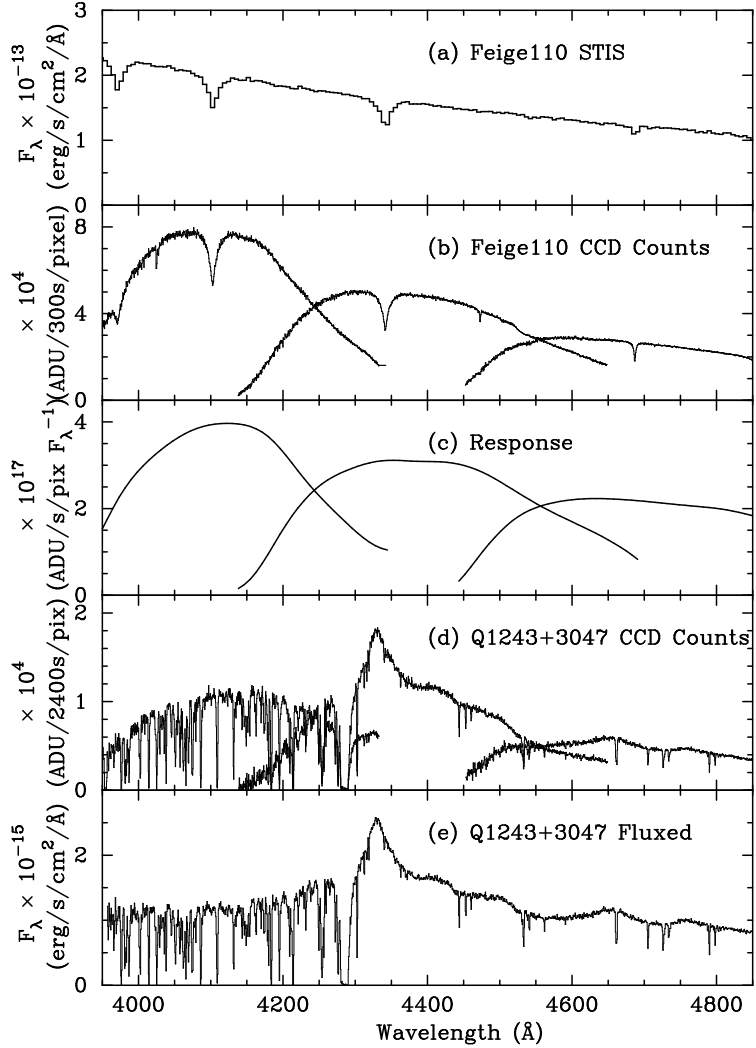}
\caption{
Steps in the flux calibration of an ESI integration of \qfive\ using
an ESI spectrum of the star Feige~110 obtained on the same night.  We
show only 3 of the 10 ESI orders. Panel (a) at the top shows the STIS
flux calibrated spectrum of star Feige 110. Panel (b) shows the raw
CCD counts from an ESI spectrum of this star. Panel (c) shows the
smoothed ratio (b)/(a) that is the response function of ESI.  Panel
(d) shows the raw counts from an integration of \qfive , and panel (e)
shows the same spectrum after relative flux calibration.  \label{X10}
}
\end{figure}

\begin{figure}
\includegraphics[angle=270,scale=0.45]{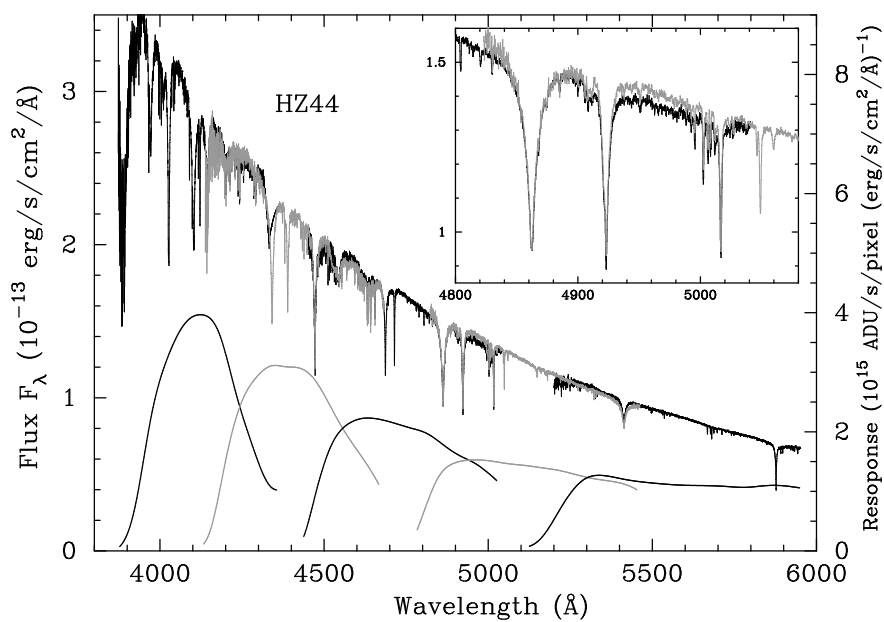}
\caption{
Demonstration of the errors in the flux calibration of an ESI spectrum of
the star HZ44. The middle traces show the ESI orders after
flux calibration using an ESI spectrum of star Feige~110.
The lower smooth curves show the response function of these ESI orders.
Here we use the usual flux calibration methods.
The insert in the upper right is an enlargement of
4800 -- 5100~\AA\ that clearly shows that adjacent orders differ.
\label{X6}  }
\end{figure}

We flux calibrate ESI spectra in the same way we do those from Kast.
In Figure \ref{X10}, we show the steps in the flux calibration of an
ESI spectrum. The ESI orders overlap in wavelength, and in Figure
\ref{X6}, we see that the orders do not always have the same flux.
These differences increase in size as we approach the end of an order.
We have not investigated the origin of these flux differences. We cut
off most of the regions where the differences exceeds 2\% (which is
the noise level in our spectrum) without leaving any gaps in the
wavelength coverage.  We do not know the size of the remaining error,
especially in regions where there was no wavelength overlap, because
we do not have redundant observations of bright stars.  Nonetheless,
the ESI spectra could have better relative flux calibration than HIRES
spectra, for several reasons. ESI has fewer orders, each of which
covers more wavelength range and a much larger portion of each order
is sampled by adjacent orders.  ESI has a fixed instrumental
configuration and our ESI spectra have much higher S/N than our
individual HIRES integrations, which may change the proportion of the
flux that is extracted.

\subsection{Errors in the Reference Spectra}

The errors in the relative flux calibration of the reference spectrum are a
fundamental limitation on how well we can apply relative flux
calibration to the HIRES spectrum. One way to explore these errors is
to compare different reference spectra.  
We will see that the differences increase with wavelength range and they
are the main source of error in our calibration of our HIRES spectra of 
\qfive .

We found that the 5 Kast spectra show two types of shape.  The two
from 1999 are similar, as are the two from 2001.  
We call the sum of the two flux
calibrated spectra from 1999 K99, and the two from 2001 K01.  KSUM 
is our name for the sum of all five spectra.  The 1997 spectrum is
similar to K01, but has much lower S/N.

These two groups, K99 and K01, differ in shape on the largest scales
across the whole \lyaf , but they do have very similar shapes across a
few hundred Angstroms after we normalize them to each other at
those wavelengths.  These differences are best seen when we smooth 
them slightly by
differing amounts to reduce differences in the spectral resolution.
The K01 spectra have up to 10\% systematically higher flux at
wavelengths $<3400$~\AA\ than do the K99 spectra.  The K01 spectra had
a few percent lower flux from 4000 -- 4300~\AA\ and higher flux across
the \lya\ emission line.  We do not know the origin of these
differences.  Possible origins include variation in atmospheric
extinction \citep{burki95} or a variation in the QSO.

We find that the ESI spectra differed from the various Kast spectra by
typically 2\% or less per Kast pixel, from 4100 -- 4400~\AA .
The differences correlate over a few Kast pixels in the \lyaf ,
with no large scale trends, except that to the red side of the \lya\ emission
line the ratio of the Kast to ESI flux increases by
approximately 5\% from 4300 -- 4400~\AA\ for all five Kast spectra.

\section{Step 2: Flux Calibrating HIRES Echelle Orders with a Reference Spectrum}

We calibrated the relative flux in a HIRES spectrum using a calibrated
reference spectrum from either Kast or ESI.  We divided a smoothed
version of the HIRES spectrum by the reference spectrum to find the
``Conversion Ratio" spectrum.  We found that the smoothed HIRES
spectrum and the reference had to have the same wavelength scale and
resolution, because the \lyaf\ absorption lines cause the flux to vary
rapidly in wavelength.  We calibrated the individual orders of each
HIRES integration using a smooth function fitted to the conversion
ratio spectra, one for each order of each integration.

\begin{figure}
\epsscale{0.5}
\plotone{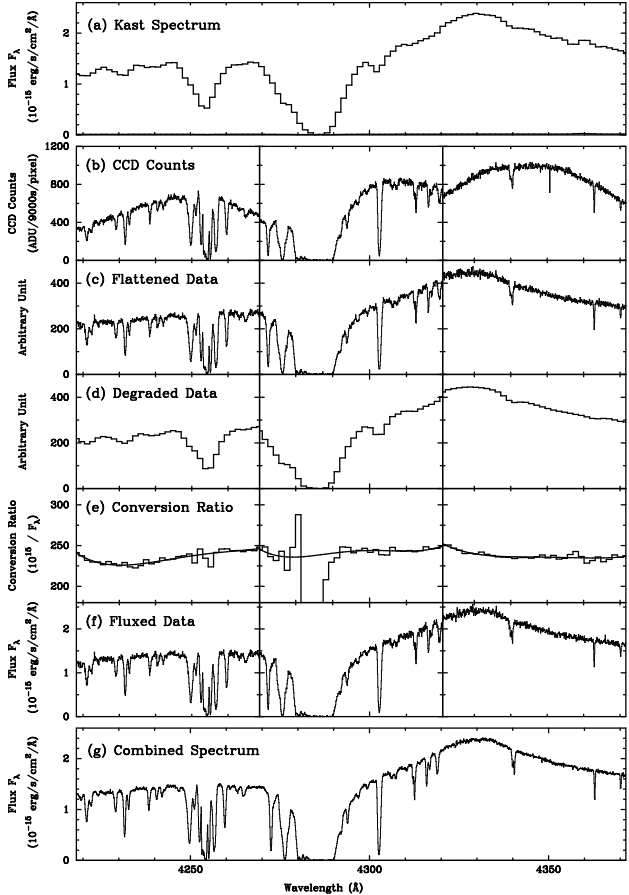}
\caption{
Illustration of the steps taken to apply relative flux to HIRES spectra of
\qfive\ using Kast spectra of the same QSO.
Panel (a) shows a flux calibrated Kast spectrum.
It has been shifted in wavelength to match the HIRES
wavelengths.  Panel (b) shows the CCD counts recorded in the three
HIRES orders that cover these wavelengths. This is a single HIRES
integration, and the orders overlap in wavelength, although we do not
show this here.  Panel (c) shows the extracted HIRES orders that have
been ``flattened" by dividing by the flat field. This is the
Flux-name.fits file that is the usual output from MAKEE.  Panel (d) is
shows the spectra from panel (c) after smoothing by a Gaussian filter
to match the spectral resolution of the Kast spectrum in panel (a).
The HIRES spectrum has been re-binned onto the Kast pixels.  Panel (e)
shows the ratio (d)/(a) that we call the conversion ratio.  It has
values at the Kast pixels, and two values at the wavelengths that
appear in two HIRES orders. The smooth curves are low order fits to
the pixels that sample the conversion ratio.  Panel (f) is (c)/(e),
the flux calibrated HIRES spectra. Notice that the jump in the HIRES
flux at the order join near 4320~\AA\ in panels (b), (c) and (d) is
detected by the conversion ratio in (e) and corrected in (f).  Panel
(g) shows the sum of 8 HIRES integrations, each of which is corrected
individually in the same manner. The order joins are not apparent.  We
do not plot most pixels in the high resolution spectra, to reduce the
file size.  \label{X17} }
\end{figure}

In Figure \ref{X17},
we illustrate the calibration of the relative flux in the HIRES orders
that we describe in the following five sub-sections.

\subsection{Wavelength Matching}
Because the wavelengths from HIRES are more accurate,
we shifted the Kast spectra onto the HIRES wavelength scale.
We measured the shifts by cross-correlation of
complete HIRES orders, and we confirmed the values by cross-correlating
individual strong lines in the \lyaf .

\begin{figure}
\includegraphics[angle=270,scale=0.45]{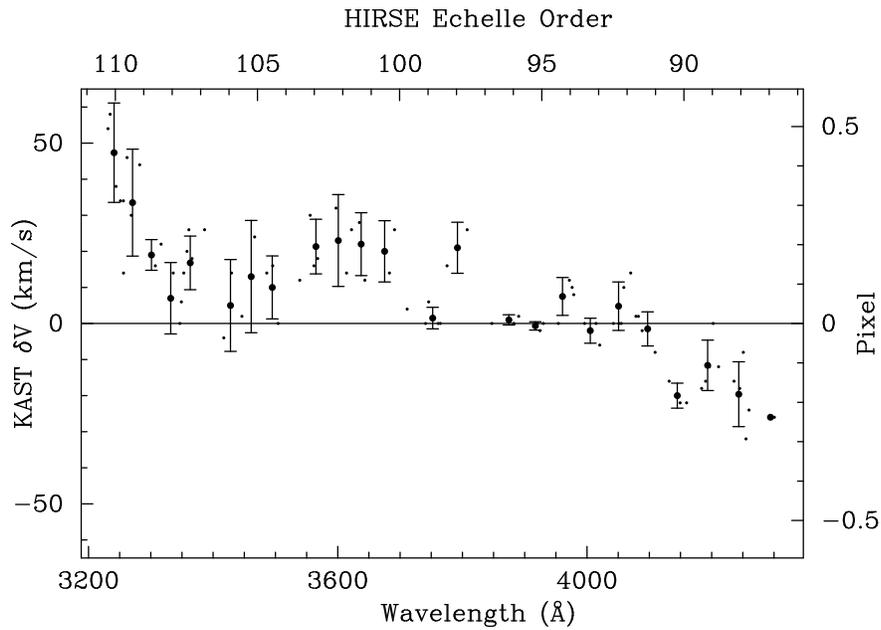}
\caption{
Shifts in the wavelength scale of a spectrum of \qfive\ from the
Kast spectrograph measured relative to a HIRES spectrum of the same QSO.
Each dot shows the shift measured by cross-correlating a region of spectrum
that contains an absorption line. The bars show means of these dots,
taken over the individual HIRES orders. 
We obtain similar shifts when we cross-correlate over complete HIRES orders.
The sampling pixel size is 107.1 \kms .
 \label{X11}  }
\end{figure}

Some of the wavelength shifts may arise because the QSO was not
exactly centered in the slit. This is a reasonable explanation for the
typical shift which was 42~\kms , or 0.4 Kast pixels, only 
17\% of the projected width of a 2 arcsecond wide slit.
These shifts should vary monotonically
along a spectrum, and some Kast spectra show this.  However, Figure
\ref{X11} shows that other spectra have more complex wavelength
shifts.  In such cases we measured the mean shift for each HIRES
order, and we fit a low order polynomial to these mean values, similar
to that used to fit the arc line wavelengths, to give a smoothly
changing wavelength scale without gaps or discontinuities.

We also placed the ESI spectra on the HIRES wavelength scale.  
This wavelength scale assignment was obtained by first smoothing 
the HIRES spectra to the approximate spectral
resolution of ESI, and then finding the velocity
shift by cross-correlation of the ESI with the HIRES integration.  
We shifted the complete ESI spectrum by +5.61 \kms , which is
7.5\% of the projected 1 arcsec slit width.  As with
the Kast spectra, this shift is larger than we would expect from the
wavelength fits to the arc calibration lines and may arise
because the QSO was not exactly centered in the slit.


\subsection{Resolution Matching}

We smooth the HIRES spectrum to match the resolution of the Kast
spectrum.  This procedure is sensitive to wavelength shifts between
the two spectra, and hence it was done after the wavelength scales
were matched.  In contrast, the wavelength scale matching is
insensitive to the spectral resolution.  We smoothed the HIRES
spectrum, and sampled it in the wavelength bins of the Kast spectrum.
We smoothed with a Gaussian function of known FWHM, truncated at
$2\sigma$, and normalized to unit area.  We smoothed by different FWHM
to find the amount of smoothing that left the smallest residuals
between the smoothed HIRES and Kast spectra.  These residuals appeared
flat across strong absorption lines, which suggests that a Gaussian is
an adequate approximation to the line spread function of the Kast
spectrum.  As with the wavelength matching, the \lyaf\ provides
additional signal for resolution matching.

The Kast spectral resolution varies from spectrum to spectrum, and it
can vary with wavelength in a spectrum, depending on where we chose to
focus.  We took the Kast resolution to be the FWHM of the smoothing
applied to the HIRES, after subtracting the initial HIRES 8~\kms\ FWHM
in quadrature.  For example, in the 1999 spectra of \qfive , the FWHM
of the Kast spectrum is near 300~\kms\ near 3300~\AA\ and from 3800 --
4300~\AA\ but it improves to 220~\kms\ near 3500~\AA , where the
measurement error is approximately 10 -- 30~\kms , depending on the
region of the \lyaf .  We smoothed the HIRES spectrum by a single mean
FWHM even when the resolution varied along the Kast spectrum.  Using
similar methods we found that the ESI spectrum had a FWHM of 63.2
$\pm$ 3.0 \kms .

\subsection{Calculating the Conversion Ratio}

We divide the smoothed HIRES spectrum by the Kast spectrum, to obtain
the ``conversion ratio" (CR) spectrum.  The \lyaf\ hinders our
calculation of the CR, because we are very sensitive to slight
remaining errors in the wavelengths and the resolution.  
When we calculate the CR, we weight the flux in the individual spectra
by their errors, and we assign an error to each pixel in the CR.
It is common to see increased error in the CR near strong
absorption lines (e.g. near 4285~\AA\ in panel (e) of Figure \ref{X17}), but 
this error does not have a systematic shape when
the wavelengths and resolution are well matched.

\subsection{Smoothing the Conversion Ratio}

We fit the CR spectra to obtain a smoothly changing function. The
original CR spectra vary from pixel to pixel because of photon
noise in the reference and HIRES spectra, especially in strong
absorption lines.  These variations can be much larger than the
expected change in the flux calibration and hence we can improve the
flux calibration by interpolating. We explored various ways of
smoothing and fitting, some in one dimension, along each order
separately, and others in two dimensions, both along and between
adjacent orders. The best choice will increase the S/N in the CR as
much as possible without changing the structure.

\begin{figure}
\includegraphics[angle=270,scale=0.45]{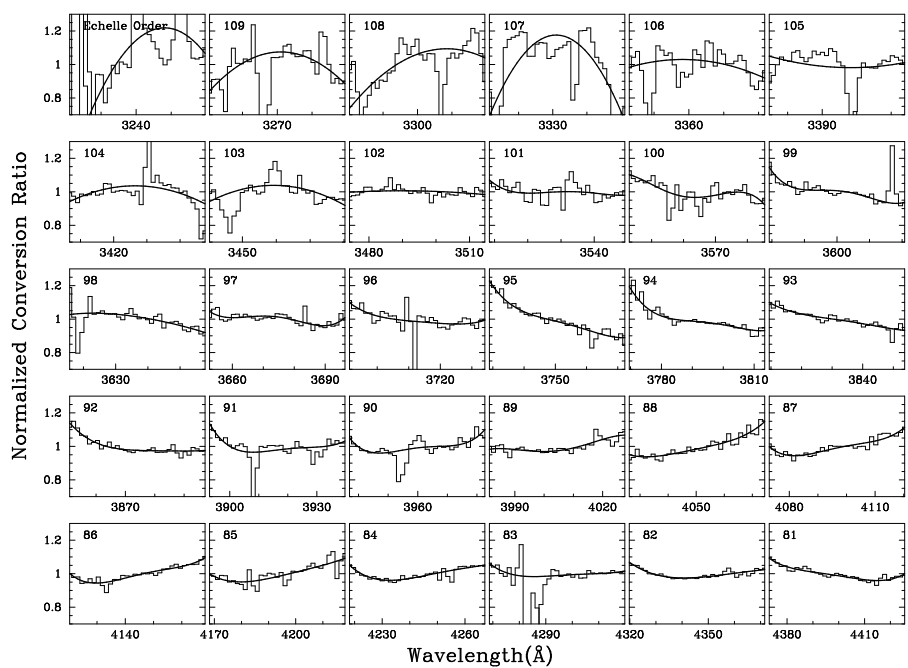}
\caption{
Conversion ratio for the 8100 second HIRES integration of \qfive\ taken April
4, 1999. The relative flux calibration uses a Kast spectrum from 2001.
The mean level of the CR in each order has been normalized for the plot.
The pixels sizes are from the Kast spectrum, and the curves show 4th order
Chebyshev polynomial fits to each order.
 \label{X21b}  }
\end{figure}

We found that a 4th order Chebyshev polynomial fit to the CR spectrum
for each HIRES order was a good choice. We choose this order by trial
and error. It leaves enough freedom to fit the shapes of the CR and
give a reduced $\chi ^2 \simeq 1$. Other fits would be appropriate in
spectra with different amount of structure and S/N.  In Figure
\ref{X21b}, we show the CR spectra for many HIRES orders.  We see
that the CR spectra for adjacent orders are very similar in shape, but
change systematically as we move across many orders. We also found
that the changes in shape from order-to-order are larger than
those between integrations for a given order, except for the effects of
photon noise in the CR.

We calculate the CR twice, in an iterative fashion, to improve the accuracy
near the ends of the HIRES orders. The first time, we stop the calculation 
before the order ends, where the filter, which we use to smooth the HIRES
spectrum to the resolution of the reference spectrum,
just touches the last pixel in the order. We can not, at that time, calculate 
the CR for the remaining pixels because we do not know the flux from beyond the
end of the order.
This flux is needed when we smooth the HIRES spectrum to match the reference 
spectrum. 
We do, however, allow the fit to the CR to extrapolate to the end of the order,
and we use that extrapolation to make the first estimate of the flux 
calibration. When we calculate the CR the second time, we 
know the HIRES flux from beyond the end of the order, because we have already
calibrated the whole HIRES spectrum.

\begin{figure}
\includegraphics[angle=270,scale=0.45]{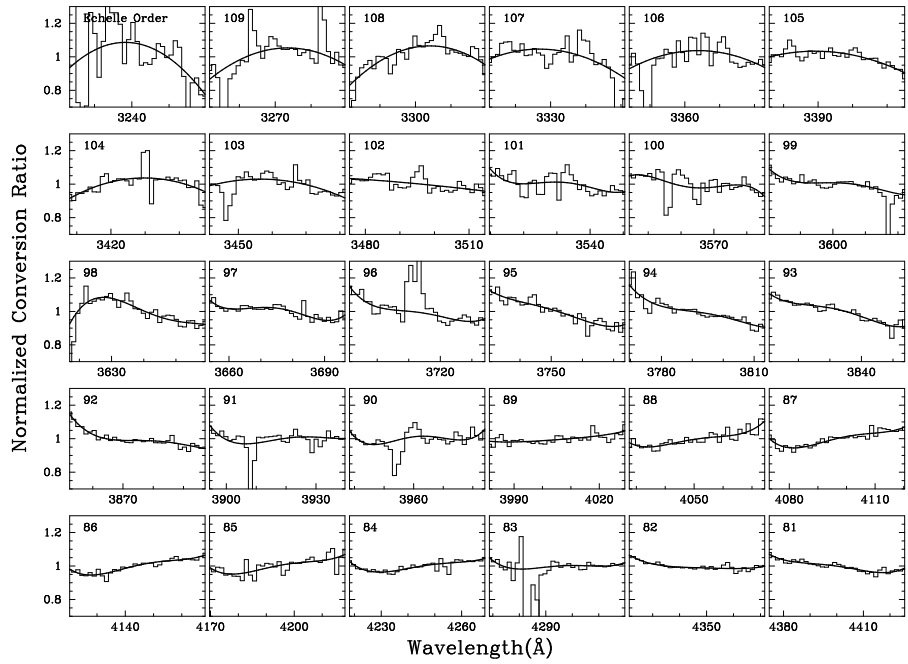}
\caption{
As Fig. \ref{X21b}, but for a 7200 second HIRES integration taken 11 months 
later on March 14, 2000. 
 \label{X21f}  }
\end{figure}

An example of the type of error that can occur in the CR at order ends
is shown in Figure \ref{X21f}. In middle panel on the left, the
CR is erroneously low in the first two pixels at the start of order 98.
Here we calculated the CR once only, and we ignored the flux from beyond the
end of the order, where there happens to be a \lyb\ absorption line.
This absorption line lowers the flux in the reference spectrum, but not in
the HIRES spectrum for that order.

The error in the relative flux calibration of a HIRES integration
depends in part on the S/N in that HIRES integration. The conversion
ratios in Figure \ref{X21f} are for a second integration of \qfive\ that
can be compared to the integration shown in Figure \ref{X21b}.  The
two integrations were calibrated using the same Kast spectrum, and
hence the differences are caused by noise in the HIRES integration,
especially at the shortest wavelengths.

\begin{figure}
\includegraphics[angle=270,scale=0.7]{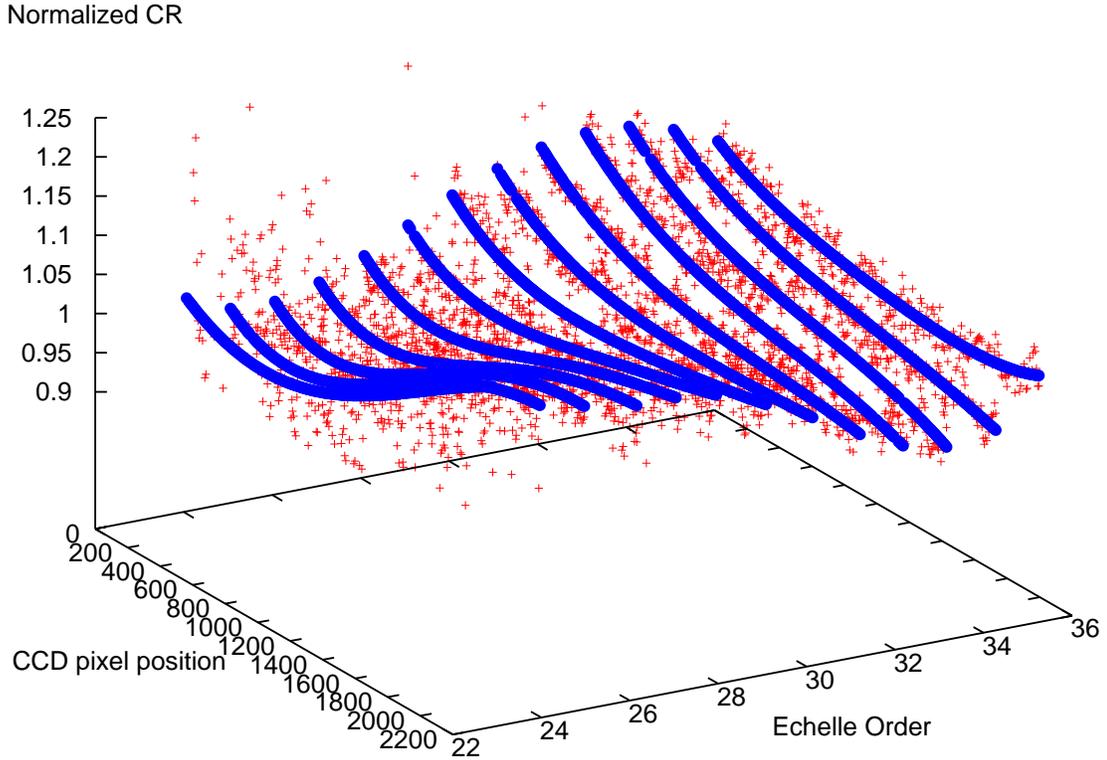}
\caption{
The conversion ratio for a HIRES integration of \qfive\
fitted with a 2-dimensional Chebyshev polynomial. Each small plus sign shows 
the CR value in a pixel of the reference spectrum from ESI.
The CR values are shown elevated above a plane that approximates
the HIRES CCD.
The axis ``CCD pixel position" is the pixel along the
direction of a HIRES order, with wavelength increasing to higher
numbers.  The axis ``Echelle Order" is encoded such that 113 - the
number given is the HIRES order. The orders are shown parallel to each
other and with equal spacing.  The vertical hight of a plus sign shows the
CR in that ESI pixel, and the other two coordinates show the location
of the HIRES pixel with a similar wavelength.  The S/N in the ESI
spectrum increases with wavelength, to the left.  The CR have been
normalized to have a mean CR $=1 $ in the middle 20\% of each HIRES
order.  Prior to this, the CR varied systematically by approximately a
factor of two.  The thick curves show the Chebyshev polynomial along
each order. These polynomials are constrained to 4th order in both the
CCD pixel and echelle order directions.  \label{twodfit} }
\end{figure}

We experimented with other ways of smoothing the CR to reduce the
effects of photon noise. The 4th order polynomials smooth in
wavelength along an order, but they do not use any information from
adjacent orders.  We made a 2-dimensional map of the CR in the
coordinate frame of the HIRES CCD detector, and we smoothed this map
in both dimensions, both using a Gaussian filter and independently by
fitting a 2D surface using Chebyshev polynomials.  The Gaussian filter
did not work well because the largest width that did not change the
shape of the CR surface had a $4 \sigma $ width of only 3 orders, not
enough to reduce the noise significantly.  In Figure \ref{twodfit}, we
show a surface fit that was an improvement over the 1D polynomial
fits, but we did not use this in \citet{kirkman03}.

The error in the CR depends on the S/N in each CR pixel, and on the
fitting or smoothing that we use. The error will correlate on the
smoothing scale. When we apply a 4th order polynomial to each
order of 40~\AA, (the correlation length is approximately 10~\AA) and
this can be seen when we compare Figures \ref{X21b} \& \ref{X21f}.
There are also strong correlations in the CR near the ends of orders,
and near strong absorption where the errors are largest.

The contribution to the error in the CR from the S/N in the reference
or HIRES spectra can be estimated from the number of pixels in the
correlation length.  There are approximately 34
pixels in a Kast spectrum per HIRES order, and hence 
8.5 per correlation length. An individual
Kast spectrum of \qfive\ has S/N 60 per pixel near 4250~\AA\ and 20
near 3250~\AA.  The CR will then have errors of at least 0.6\%
near 4250~\AA\ and 1.7\% near 3250~\AA.

\subsection{Applying the Conversion Ratio}

We divide the original HIRES integration, with full spectral
resolution and pixels, by the 1D fits to the CR to obtain the desired
flux calibrated HIRES spectrum.  Since we have not merged the orders,
the wavelength overlaps remain.  The resulting HIRES integration now
has relative flux calibration on each order.

\section{Step 3: Combine the HIRES Orders}

The final step is to combine the fluxed HIRES orders into a single
spectrum.  We add the HIRES integrations that have very similar
wavelength coverage, order by order, and then merge the orders. We choose
this sequence because it facilitates checks of the relative flux
calibration.  We compare the flux in each order in multiple
integrations before we sum the integrations. After they are summed,
the enhanced S/N makes it easier to see errors in the flux
calibration in the wavelength overlap between orders.

First, we place all orders from all HIRES integrations on a single
wavelength scale, so that pixels from orders that overlap in
wavelength sample exactly the same wavelengths.  We choose a scale
with a constant velocity increment of 2.1 \kms\ per pixel, equivalent
to constant log wavelength increment.  This choice has two advantages
over constant wavelength: in velocity units, the spectral resolution
of the echelle does not vary significantly with wavelength, and we use
velocity units to describe the widths of absorption line systems.

We simultaneously correct errors in the wavelength scales, using
information from the cross-correlation of different integrations as
discussed in (\S \ref{hireswave}). We do this correction
at the same time because we want to re-bin the HIRES spectra only once.

We calculate the mean flux in each pixel, from all integrations, after
rejecting large deviations that we identify by evaluating the $\chi
^2$ statistic for each pixel.  We ignore the flux in a pixel from an
individual integration if it is $> m\sigma _i$ away from the weighted
mean for all integrations, where $\sigma _i $ are the errors on the
fluxes in the individual integrations.  The value of $m$ was
determined iteratively to remove most pixels with conspicuously
deviant flux values. For \qfive , we use $m=2$.  The algorithm
iterates, and re-derives the $\chi ^2$ after rejecting the flux value
from one integration. If all integrations are $> m \sigma$ away from
the weighted mean, we examine the errors.  If the errors in some of
the integrations have data, shown by non-zero errors returned by
MAKEE, we use the flux from the integration with the smallest error as
the mean.  If, however, there is no data in a particular pixel, for
example because of a known CCD defect, then we use the weighted mean
of the two pixels on either side, again using only those integrations
that are within $m \sigma$ of the weighted mean for that wavelength.
If none of the above criterion are met, we use the average of the
fluxes in all integrations.

Finally, we verify that adjacent orders show similar flux levels where
they overlap, and we used a weighted mean to combine the orders in
these regions, producing a single flux calibrated spectrum.  In Figure
\ref{f344thord}, we show order overlap for spectra of star Feige
34. This spectrum can be compared to the spectra in Figures 
\ref{X16} that show
the flux calibration using usual methods. The ratio of the flux in
adjacent orders shows most variation near 3250~\AA\ : 0.92 -- 1.08\%,
and decreases to 0.98 -- 1.02 near 4300~\AA , where the signal to
noise ratio is highest in the Kast spectrum that we used as a reference.
The calibration of Feige 34 used the method and algorithm that we
developed and applied to our spectra of \qfive , with a few
exceptions.  The star does not have \lyaf\ lines, and hence we matched
the wavelength scale of the Kast and HIRES spectra using a single
shift for the whole Kast spectrum. We fitted the CR with one 4th order
polynomial per order.  A 6th order fit would work better near
4300~\AA\ where the signal to noise ratio is high.

We consistently found, from many spectra, that the remaining difference 
between between the orders was 
largest at the end of an order, where the CR is less well known. Hence,
we tapered the flux from each order using a weighting function that
declined linearly from one, where the overlap begins, to zero, where the order
ends.

\section{Comparison of Spectra of \qfive\ Calibrated with Different 
Reference Spectra}

We made several different calibrations of the HIRES spectra of \qfive\ using
different low resolution spectra to convey the flux information.  We
label these HIRES spectra by the spectra that we used for
the flux calibration: hence, HK99 is the HIRES spectrum of \qfive\
calibrated using the Kast K99 spectrum of \qfive, HKSUM was calibrated
using the mean of all five Kast spectra and HESI was calibrated using the
ESI spectrum of \qfive .

We also made a spectrum, HH, which we calibrated using a HIRES spectrum
of a standard star. This HH calibration is not typical of the accuracy that 
we usually obtain, but rather it is the best that we could obtain with our
spectra. We made many calibrations using different HIRES integrations of 
standard stars, and we show the calibration that had the smoothest order 
joins. This HH calibration contrasts with the calibrations of the stars that
we showed in Fig. \ref{X16} that did not work as well.

\begin{figure}
\epsscale{0.7}
\plotone{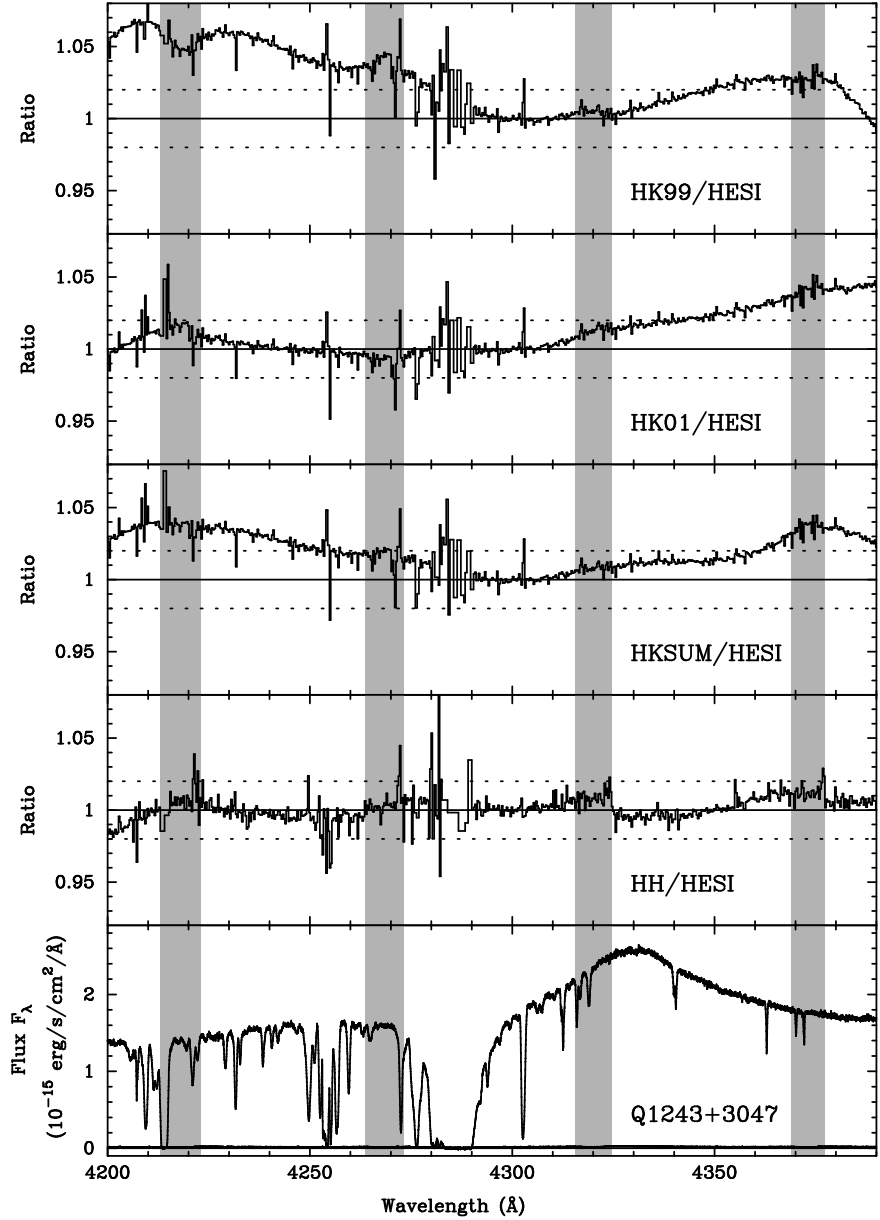}
\caption{
Ratios of the flux in different summations of the 8 HIRES integrations of
\qfive\ that we have calibrated using different spectra.
HK99, HK01 and HKSUM were all calibrated with Kast spectra, while HESI
and HH are HIRES spectra calibrated using ESI spectra and a HIRES
spectrum of a flux standard.  Each of the top 4 panels shows the ratio
of two HIRES spectra, each one of which looks similar to that shown in
the bottom panel.  The vertical bands show the wavelengths where the
orders overlap.  We do not plot most pixels, to reduce the file size.
If we had plotted all pixels, the noise near the few strongest absorption
lines would be much more conspicuous, and in each 10~\AA\ interval
we would see 1 -- 20 fluctuations of 1 -- 2\%.
We also do not plot pixels that have negative flux, because of the 
random noise in the sky subtraction.  
\label{X15} }
\end{figure}

In Figure \ref{X15}, we show the ratios of HIRES spectra calibrated in
these different ways.  The top panel shows HK99 spectra divided by
HESI.  There is a 7\% difference across the wavelength region shown.
The HK01 differs from HESI by 5\% at most, while the HKSUM
calibration differs from HESI by 4\%. The calibrations HH and HESI
differ the least -- only 3\% -- but show the largest jumps at order
joins, e.g. near 4325~\AA .  
The four gray bands of Figure \ref{X15} show the wavelength 
overlaps between the HIRES
orders. We show three complete orders and two partial ones.

We show the wavelength region centered on the
strong \lya\ lines that we have use to measure the H~I column density
that contributes to a D/H measurement. This \lya\ line has damping
wings that absorb approximately 1\% of the flux near 4233 and
4340~\AA\ and hence accurate flux calibration of this region helps us
measure the column density, although most information comes from
20~\AA\ on either side of the line center.

The differences between the calibrations have three origins.
The largest differences, which are on the largest scales, 
come from the differences between the K99, K01 and ESI spectra. 
Other differences, especially near the order joins, are related to the 
fitting of the CR, and are sensitive to the signal to noise ratio in the Kast 
and ESI spectra. A third type of difference occurs from pixel to pixel,
and comes from the rejection of deviant pixels from among the 8 HIRES 
integrations. The CR fits are smooth curves,
and the ratios would also be smooth were it not for differences in the
pixels which are rejected when we took the sum of the 8 integrations.
The numerous small 1 -- 2\% deviations arise when a pixel is not used
from one integration or another. The size of these deviations is
given by the signal to noise ratio of the individual integrations. 
The noise increases
in the strong absorption lines where we are dividing two HIRES
spectra, each of which has low signal to noise noise. 
Trends that are seen in more than one panel may come from differences
of the ESI spectrum from the others.  

We do not know which of the spectra is the more accurate.  
All of the spectra used in Figure \ref{X15} were calibrated with CR spectra 
that were fit order by order. We found that the 2-dimensional fits to the CR,
like that shown in Figure \ref{twodfit}, slightly reduced the deviations 
near the order joins.

\section{Discussion of the Accuracy of the Flux Calibration}

Many factors contribute to the errors in the relative flux calibration,
including:
\begin{itemize}
\item Errors in the flux reported for the standard star.
\item Errors in our spectra of the standard star and reference. Common
error sources include
extinction and absorption in the Earth's atmosphere, slit losses that depend 
on wavelength, a dichroic in the 
spectrograph, variation of the target, variation with wavelength of
the proportion of the flux extracted, and the S/N of the spectra.
\item Errors calculating the response function that we use to calibrate the
reference spectrum. 
Errors occur matching the resolution and wavelength scales of our spectrum of 
the standard to the published flux information.
Such errors are especially conspicuous near absorption lines in the standard.
\item Errors in the preparation of the HIRES spectrum, including
the bias subtraction, flat field division and extraction.
\item Errors applying the flux calibration to HIRES spectrum, including
wavelength shifts, resolution differences, the signal to noise ratio of the HIRES 
spectrum and fitting the CR.
\end{itemize}
We have found that many of these factors can produce 1 -- 10\% errors in flux
calibration, but it is difficult to assign typical values for these errors.

The size of the error in the relative flux calibration depends on the
wavelength and the wavelength range.  We do not include errors from the CR
in the usual error array because the CR errors are 
correlated over many pixels.
In this paper we have concentrated on scales of a few orders, or 
approximately 120~\AA\ that are most relevant to our work on D/H. We
have paid much less attention to the relative flux calibration on larger scales
that are dominated by a different set of factors, such as the extinction
at the time of observation.  We expect errors due to extinction to
monotonically increase with scale in our reference spectrum.

\begin{figure}
\epsscale{0.8}
\plotone{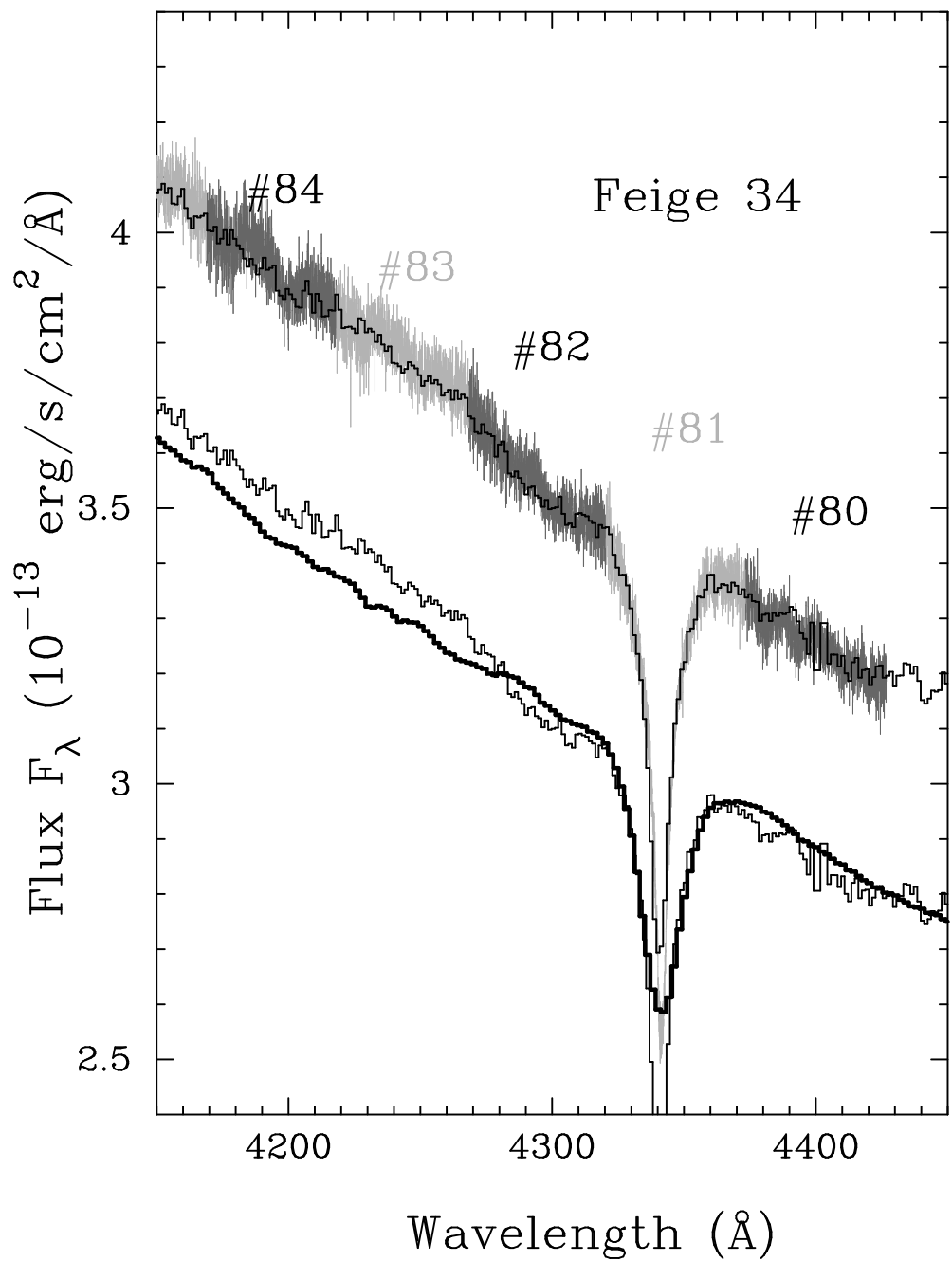}
\caption{
Spectra of star Feige 34 with relative flux calibration.
The lower two spectra are from STIS (heavy line) and Kast (thin line),
both from Fig. \ref{X23}.
The upper two spectra, displaced vertically by the same amount for clarity, 
are the same Kast reference
spectrum and five and a half orders of a HIRES spectrum 
(faint trace with many pixels, darker in even numbered orders).
\label{f344thord} }
\end{figure}

An indication of the accuracy we have attained in relative flux
calibration is given in Figure \ref{f344thord}. We compare three 
spectra of the flux standard star Feige 34: a HIRES spectrum that we
flux calibrated using a Kast spectrum as the reference, the Kast
reference spectrum, and STIS a spectrum.
The Kast and STIS spectra are those shown in Figure \ref{X23}.
The STIS spectrum was not used in the calibration, except to provide the
normalization of the Kast spectrum across the range 3200 -- 4450~\AA .
We used a one-dimensional 4th order polynomial to fit the CR.
In the wavelength region shown, the HIRES spectrum differs from 
its reference by at most 2.5\%, except near the absorption line,
and typically $<1$\%. 
At wavelengths 3400 -- 3800~\AA , where the signal to noise ratio is lower, the
differences are twice as large.  
The differences correlate on scales $> 5$~\AA .
The flux in different HIRES orders joins smoothly, with no unusual structure.
The remaining differences of the HIRES and STIS spectra come from the
deviation of the Kast spectrum from the STIS, shown in Figure \ref{f344thord}.
This comparison demonstrates that the method can give errors of
$<1$\% in the relative flux over approximately 200~\AA .  
For Feige 34 the accuracy of the flux calibration was limited by the 
accuracy of the reference, and not by the method itself.
\section{SUMMARY}

We found that the distribution of the signal recorded in HIRES
integrations differs from integration to integration. We 
do not have a complete explanation for this behavior, although
varying vignetting and inadequate extraction may be involved.
We found that these differences persist
even when the instrument is apparently unchanged. These changes mean
that the usual methods of flux calibration are inadequate.

The methods we have described for applying relative flux calibration
to a HIRES spectrum use three spectra: the HIRES spectrum of the target that
we wish to calibrate, and Kast spectra of both the target and a flux standard
star. We use the latter to get the Kast response and calibrate the
Kast spectrum of the target that we name the reference spectrum.  We use 
reference spectrum to impose flux calibration on the HIRES target spectrum.

This has method has three advantages. First, we
can calibrate HIRES when normal calibrations using standard stars
observed with HIRES alone are inadequate.  Second, we can correct many
types of error in the HIRES spectrum, including those from varying
vignetting and inadequate extraction.  Finally, we can obtain all the
calibration spectra at a different time and on a different telescope.

The error in relative flux calibration, and the solution that we
describe, could apply to any spectrum with inadequate relative flux
calibration, whether from an unstable spectrograph or from inadequate
extraction. Vignetting could vary in any spectrograph that was unstable.
Instability could involve an optical misalignment, as with HIRES.
Variable vignetting would be harder to recognize in a
first order spectrum because we expect the largest changes near the
largest field angles, but there is only one spectrum to show this
change, and
flux calibration is often harder near the ends of a single spectrum, for other
reasons.

\section{Acknowledgments}

This work was funded in part by grant NAG5-9224 and by NSF grant
AST-9900842 and AST-0098731.  The spectra were obtained from the Lick
and the W.M. Keck observatories. The W.M. Keck observatory is managed by a
partnership among the University of California, Caltech and NASA.  We
are grateful to Steve Vogt, the PI for the Keck HIRES instrument that
enabled our work on D/H, for many invaluable discussions, along with
Tom Barlow, who developed the MAKEE extraction software package.  We also
thank Wako Aoki and Kunio Noguchi for discussions of the HDS echelle 
spectrograph, and the W.M. Keck Observatory staff and the Lick Observatory 
staff.

\appendix

\section{Appendix A: Choice of Standard Star}
\label{appa}

\begin{figure}
\includegraphics[angle=270,scale=0.45]{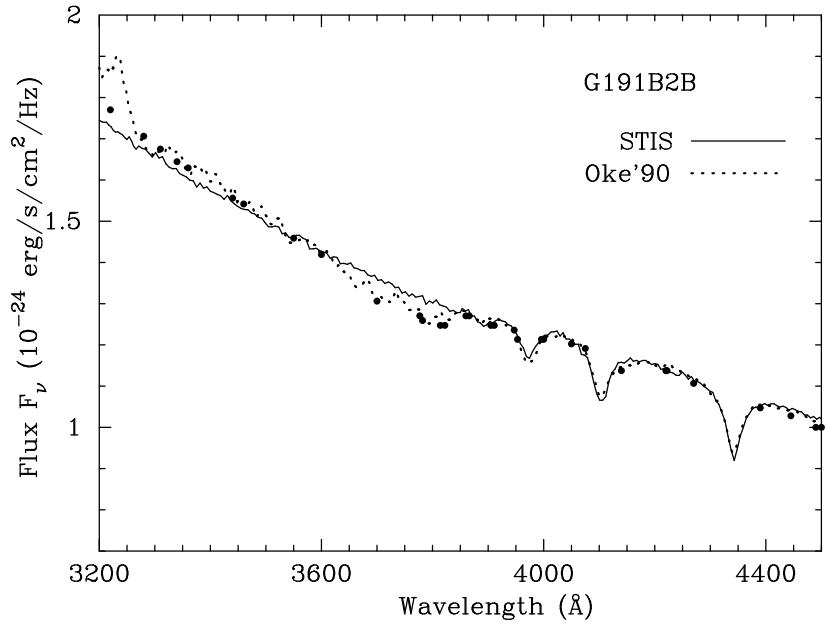}
\caption{
Flux calibrated spectra of the star G191-B2B.
The continuous, wobbly line is a STIS spectrum from \citet{bohlin00}.
The dotted line and points are from Oke (1990).
 \label{X4}  }
\end{figure}

We used the calibrations of the flux in stars based on
STIS spectra \citep*{bohlin01}.
In Figure \ref{X4} we see that the STIS spectra do not show the
wiggles at 3200-3850~\AA\ that are present in the \citet*{oke90}
spectrum of G191 B2B. This and other Oke spectra are
widely used by default in the reduction packages IRAF and MIDAS.
The differences between the \citet*{oke90}
and STIS spectra can reach several 
percent, larger than our random photon noise. Based on the lack of 
features in the spectrum and the STIS data quality,
we preferred the following stars for UV flux calibration near 3200~\AA :
G191 B2B, BD+28 4211, Feige~110, and Feige~34. 
For G191 B2B we have the additional choice of using a model spectrum 
given by \citet{bohlin01}. This model spectrum fits
their STIS spectra to within 0.7~\% in the continuum \citep*{bohlin00}, and
it simplifies flux calibration because it has full resolution.

We used the entire STIS spectrum for flux calibration,
including the broad Balmer absorption lines.  
The \citet*{oke90}
paper provides AB magnitudes at discrete continuum 
points in 5-50~\AA\ intervals. These points skip the Balmer absorption lines,
but we can not do this, because we then have insufficient information to
calibrate several echelle orders, each of which is only 30-60~\AA\ long.

\begin{figure}
\includegraphics[angle=270,scale=0.45]{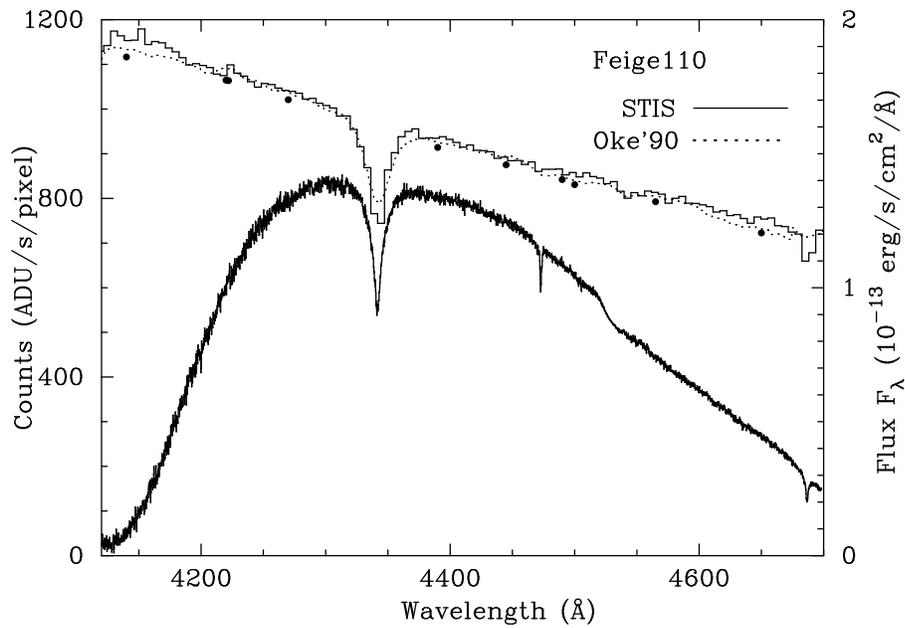}
\caption{
Spectra of the flux standard star Feige 110.
The lowest trace shows the signal recorded in one ESI order.
The dotted line shows the flux reported by Oke (1990)
and the points show the flux values that he recommended to minimize
sensitivity to spectral resolution. The STIS spectrum from
\citep*{bohlin01} is shown by the continuous
trace comprising pixels that are easy to see on the plot. 
 \label{X5}         }
\end{figure}

Our use of the flux calibration information near the Balmer lines can help us
avoid significant errors.
In Figure \ref{X5} we show a spectrum from ESI echellette order 15 that has its 
sensitivity peak around 4350~\AA\ that coincides with Balmer $\gamma $ line, 
4341.68~\AA .  \qfive\ \citep{kirkman03} happens to have its \lya\ 
emission at 4330~\AA . In an early flux calibration of this order, poor 
interpolation across this Balmer line had led
us to make an 8\% calibration error that was three times the random error.

\section{Appendix B: HIRES Image Rotator}
\label{rotator}

The vignetting in HIRES depends on whether or not the image rotator is
used, and on the mode in which the rotator is used.

HIRES is fixed to the Nasmyth platform of the Keck-I telescope with its
slit approximately parallel to the horizon.
When we look at the image of the sky on the HIRES slit, we see that the
vertical direction in this
image rotates at a rate given by the elevation (EL) of the telescope.
This is because telescope is rotating in EL 
while HIRES is fixed.
If the telescope is pointing at the horizon, and looking at an 
arrow in the sky that is pointing towards the zenith, the image of 
this arrow on the slit plane is also pointing towards the vertical, 
which is perpendicular to the length of the slit.
As the telescope moves to higher EL, the arrow rotates until it is 
aligned along the slit when the telescope is pointing to the zenith.

The HIRES image rotator allows us to rotate the image of the sky on the
slit plane in any way we like.
We installed an image rotator in HIRES in late 1996. This is a large
quartz prism that sits in front of the HIRES slit. The light from the
Keck-I telescope tertiary mirror 
undergoes three total internal reflections in the prism before 
coming to a focus on the slit plane. 
The prism can rotate continuously in either direction
about the axis of the beam that converges on the center of the slit.
The prism is aligned so that the image of a star on the center of the
slit moves by under 0.5 arc seconds when the prism is moved in or
out of the beam, and when the prism is rotated. The prism can be 
spun rapidly to demonstrate this alignment. The prism does not
vignette any of the beam that lands within 60 arc seconds of 
the center of the slit.

The image rotator has two main modes of operation: Position Angle and
Vertical Angle. The position angle mode is used when we wish to keep two
stars in the slit, where as the vertical angle mode is used to keep the
vertical direction in the sky parallel with the slit, as a surrogate for
an atmospheric dispersion compensator. 
When the image rotator was used in Vertical mode, the position angle along 
the slit is the parallactic angle, and this varies as we track a target.
The parallactic angle is measured at the target, from the North Celestial 
Pole to the Zenith, in the direction from North via East. 
The parallactic angle is fixed for a given elevation and azimuth 
in the sky, but it changes when we track a target across the sky.

HIRES spectra are hard to flux calibrate in part because the 
vignetting can change by 10\% from spectrum to spectrum.
The vignetting changes because there is a known misalignment 
between the beam coming from tertiary mirror and the HIRES optical axis.
When HIRES was installed, the center of the Keck telescope pupil 
was measured
to be approximately 9~mm away from the collimator center, which
corresponds to a beam misalignment of 7.4 arcminute. 
If a star is held at one position on the HIRES slit, the axis of 
the beam entering HIRES will rotate around the HIRES optical 
axis at a rate given by any change in the position angle of the 
sky image on the slit. If the position angle moves through 360 
degrees, the axis of the beam entering HIRES
follows the surface of a cone with an apex angle of 
approximately 14.8 arcminutes.
Steve Vogt used ray tracing to find that this rotation causes 
the vignetting to vary by approximately 10\%, depending on the angle.
The vignetting occurs due to a dewar which forms a central obstruction
in the beam near the camera's prime focus.


\clearpage

\clearpage

\clearpage

\clearpage

\clearpage

\clearpage

\clearpage

\clearpage

\clearpage

\clearpage

\clearpage

\clearpage

\clearpage

\clearpage

\clearpage



\clearpage

\clearpage


\end{document}